# Effect of Robot-Kinematic Activity on CLASS And TUG-K Scores of Non-Science Majors

*Stanley Sobolewski, Brandon Vought, Kathryn Hagood*
*Department of Physics, Indiana University of Pennsylvania*

**Abstract**

There were two objectives. The first was to determine if a real-time graphing activity will influence the subject's interpretation of kinematic graphs, as measured by the TUG-K. The second was to determine if attitude, as measured by CLASS, has a connection with graphical interpretation. The graphing activity did not affect graphical interpretation; this might be due to the short time spent on the treatment. However, there was a connection between attitude and graphical interpretation. The subjects of this study were non-science majors attending a medium-sized state-owned university.

*This work was supported by an Indiana University of Pennsylvania Senate grant*

**Introduction**

It has been shown that attitude plays a critical role in learning. (Cahill et al., 2018) This is especially true in the sciences. (Osborne, Simon, & Collins, 2003) Most of the concern is about students adopting a career in STEM; however, our position is that the attitude and knowledge about science are essential for the non-scientist as well. The lack of scientific knowledge can be astonishing sometimes. A British study (Sturgis & Allum, 2004) finds that "about 10% of US citizens correctly defined science as having to do with the concepts of controlled experimentation, theory and systematic variation" and "only 34% of the British public knew that the earth goes around the sun once per year."

There are two related research questions for this project. 1) Will a student-centered, technology-based activity influence the ability of novice physics students to interpret kinematic graphs? More importantly 2) Does attitude about learning physics differ between physics students and non- science students

With the non-STEM student in general, what is of interest is not so much the ability to solve physics problems, but to gain an appreciation of the nature of science. While this is true of the general population, it is especially of concern in elementary school teachers. A study that compares STEM to non-STEM and Elementary Education majors (Michaluk, Stoiko, Stewart, & Stewart, 2018) suggests that pre-service elementary teachers and non-STEM major students have negative attitudes about mathematics and science relative to students pursuing STEM degrees.

An essential skill for an educated individual is the interpretation of graphical information. This means more than just reading numbers from a bar graph but finding rates of change and mean values from traditional Cartesian representation. The physics education community is replete with studies showing the efficacy of real-time graphing. (Araujo, Veit, & Moreira, 2008) A classic investigation of this type has students walk or run, so their motion matches a displayed position vs. time or speed vs. time graph. This activity is widespread in physics courses for physical science majors; it is often used in high school physics as well. (Beichner) However, this technique is seldom used in courses for non-science majors.

Some non-science students often have a poor opinion of physics, sometimes feeling "I cannot do it," "it is too hard," or "it is not important." (Fortner, 1993) (Osborne et al., 2003) *The primary goal of this study is to find if hands-on, interactive, enjoyable, and hopefully, entertaining laboratory activity will influence the non-science student's opinion of physics. It is expected that there will be a positive impact on the subject's attitude toward science.* A second goal will be to find if the understanding of the subject of the Newtonian description of motion will improve. Due to the increased interest of robotics in education, there are quite a few robotics kits that are easily programmed through a computer interface. (Mitnik, Recabarren, Nussbaum, & Soto, 2009) When micro-robots (MR) were used in an engineering course, Yamanishi et al. (Yamanishi, Sugihara, Ohkuma, & Uosaki, 2015) found heightened interest in a programming language. Results from (Mitnik et al., 2009) show that students using robotic activity achieve a significant increase in their graph interpreting skills. Moreover, when compared with a similar computer-simulated activity, it proved to be almost twice as effective. The use of novel technology in instruction has been shown to increase interest in the subject matter, implementing the technology.

Instruments to be used include the Test of Understanding Graphs – Kinematics (Lasry, Rosenfield, Dedic, Dahan, & Reshef, 2011) and the Colorado Learning Attitudes About Science Survey (CLASS). (Semsar, Knight, Birol, & Smith, 2011) Both instruments have well-established histories in the physics education community and have a long history of being a valid instrument.

Colorado Learning Attitudes About Science Survey

It has been found that student achievement expectations and academic self-concept were more significant predictors then prior achievement. The CLASS was developed to measure students' beliefs about physics and physics learning. (Adams et al., 2006) Subjects are presented with thirty statements; they mark if they agree or disagree with the statement. An example statement is: "When I solve a physics problem, I locate an equation that uses the variables given in the problem and plug in the values." The subject's responses are compared to those given by professional physicists. The results from the assessment are typically presented on a chart comparing the percentage of favorable to unfavorable responses. During development, a predetermined set of latent variables had been developed. The intent of the instrument was to identify the prevalence of these categories. These are Independence, Coherence, Concepts, Reality world view, Reality personal view, Math, Effort, and Skepticism.

Test of Understanding Graphs in Kinematics (TUG-K)

The Test of Understanding Graphs in Kinematics has been designed to measure the conceptual understating of kinematic graphs. It is a multiple-choice assessment developed in the early 1990s. The distractors contain common misconceptions displayed by most novice physics students. The connection between the score on the TUG-K and logical thinking has been established by Bektasli. (Bektasli & White, 2012). Results from the Middle Grades Integrated Process Skill Test and the TUGK show students with a high logical thinking skill perform better on kinematics graph interpretation tasks related to slope than a student with low logical thinking skills.

**Epistemological Framework**

The development of a framework to understand the nature of students' misunderstanding of physics is a fundamental task in physics education research. It has become widely accepted as truth, among those who follow or participate in science education research, that students come to physics courses with conceptions about the world that differ from the physicists, and that this misconception should be addressed in instruction. The working hypothesis is that for non-science majors, a positive attitude will be related to an increase in making correct predictions regarding physics problems. In this context, the understanding of graphing as measured by the TUG-K. While it is appropriate to compare the conceptions of undergraduate physics students to expert physicists, this goal may or may not be suitable for non-scientists.

There are at least two types of misunderstandings. One concerns the fundamental nature of science. Students have difficulty with the concept of certainty in science; they do not appreciate the idea that science is evolving. (Beck-Winchatz & Parra, 2013) While this is certainly a concern to physics instructors, a more immediate concern of issues of misconceptions of motion and force. This misconception in physics students is well documented. Motion implies force, acceleration as the second rate of change of displacement; these are difficult to grasp. (Halloun & Hestenes, 1985) An aside note is that these two categories of misunderstandings seem to contradict each other. The certainly of kinematics, as well as Newtonian dynamics, seems to contradict the notion of change and uncertainty. The various models of the universe that have been developed are not usually appreciated by the novice physics student. When considering the misunderstandings motion and force, it has been found that physics students initially have ideas about physics that are comparatively similar to physics faculty; these ideas were consistent during the university experience. The "expert like views about physics, as measured by CLASS, is largely a pre-existing trait of students who choose to be a physics major rather than something developed at the university." (Gire, Jones, & Price, 2009)

While this is important to know, this research should be compared to the studies of the non-science student. It may be true that both scientists and non-scientist attempt to create mental models (Brewer, 2001). However, when considering physics instruction, should the student who has no interest in physics and does not plan to study science receive the same education as the future physicist or engineer?

According to Hendrickson, attitudes are the best predictor for the estimation of students'success    (Hendrickson, 1997). Student's reception of learning strategies must be planned and organized so that students develop a positive attitude. A student's attitude toward a subject can have a significant impact on the perception of the content. (Guido, 2018)

# Research Design

## Assessment Instruments

The population to be considered for the study are non-science students. These students would not have taken high school physics, nor college physics so they will not have been exposed to this type of experience. Physical Science class is a linked lecture-lab course in physics for non-science majors. There is a wide distribution of majors in this course, ranging from Math and Computer Science to Fashion Merchandising and Hotel and Restaurant Management. Two lab sections taught by two different instructors were selected for the experimental group. One concern is the sheer volume of items in each of these instruments Assignment to the treatment group or control group will be based upon convenience; selected lab sections will be used as a treatment group. Students in both groups took a pre-test consisting of questions from the CLASS and the TUG-K. Students in the treatment group took a similar posttest from the same assessments to determine a change in conceptual understanding as well as attitude.

During the actual lab class, the students will perform the investigation using the system developed. Students will construct position-time graphs and then observe the robot moving according to the graphs they have drawn. Following the laboratory activity, the students in the treatment group will be asked to retake the TUG-K and the CLASS, to determine a post-test score for each student on both instruments.

The objective of the study was to determine if the student-centered activity would have an impact on student attitude as measured by the CLASS and understanding of graphing as measured by the TUG-K. The control activity used photogates to measure the instantaneous speed of a cart at various points down the ramp. Students would record the position and distance traveled and time interval of the travel of the cart at eight points on the ramp. From this description, the students would draw a position vs. time, speed vs. time, and acceleration vs. time graph.

All students at this institution are required to take two courses in science, one of which must have a lab. The subjects of the study were enrolled in a non-science major's conceptual physics class that included a lab. The course uses Paul Hewitt's Conceptual Physics for a textbook. Students enrolled in this class came from a wide assortment of majors; more than half of the students listed one of the following as their major: Criminology, Fashion Merchandising, Interior Design, Middle-Level Education, Accounting, Communications Media, and Marketing. The course had a lecture component that met three hours per week and a lab section of 25 students that met for two hours once per week. Five lab sections were selected as a control group, and four performed the treatment activity.

## The Treatment

The lab portion of this course was used as a venue for experimentation due to the small size of the lab class when compared to the larger lecture sections. Smaller lab course sections also allowed for more variation between groups of subjects. The objective of the lab portion of this course is to familiarize the non-science student with basic ideas in physics. The lab did not require any preparatory activity, but it was discussed in the lecture so the students new the kind of action they would be performing. When the student enters the lab class with pre-preparation, that does have an effect on attitude. (Van De Heyde & Siebrits, 2018)

When learning, instantaneous feedback is most important. One of the goals of the instructional design is to have students make sketches of kinematic graphs and then instantly observed in the drawing matched the motion they imagined. Subjects were given a verbal description of movement. Typically, this was with a constant zero or non-zero acceleration. A sample motion description might be: "an ant crawls along the table with a constant speed so that he travels one-half meter in 10 seconds."

**Results**

The subjects were asked to respond to a 57-question online questionnaire. This questionnaire was composed of two instruments combined: the TUG-K and the CLASS. They were analyzed separately for the sake of this study.

Student's comments were solicited, most were positive:

> "The scribbler was highly underwhelming. We are told we are to work with robots in lab one day and they hardly work." Fun, The robot was cool, , The second graph of the robot activity was kind of confusing. Most of the labs do not help students learn concepts. We just follow the directions, ask for help when we get stuck, and hope we did it correctly. i (*sic*) thought it was interesting and i had never used that before"

It was hypothesized that there would be a difference in the CLASS measured attitudes in the non-science majors when compared to the physics majors enrolled in calculus-based physics. As measured by other researchers, it seems the factors identified in this study are like those identified in other studies that use the CLASS (Perkins, Adams, Pollock, Finkelstein, & Wieman, 2005) However, while Perkins was attempting to identify latent variables, we started by accepting the latent variables identified in earlier studies. We are going to determine the weights or coefficients that are of the previously identified components as they apply to the TUG-K scores in this population. Therefore, a principal component analysis was used.

The score on the TUK-K was chosen as the dependent variable, and our working hypothesis was how does attitude influence understanding.

To start this process, we perform a linear regression with the change in the TUG-K as the dependent variable. In general, the result we would get is a relationship in the following form

$$x = \beta_1 y_1 + \beta_2 y_2 + \beta_3 y_3 \ldots$$

Where x is the dependent variable, the y's are the independent variables, and the beta corresponds to the weight each of the independent variables has on the dependent variable. Since the treatment time was only one two-hour class period and the assessment was administered on-line one week later, a 90% level of confidence will be used, so we will consider p-values of 0.10 and lower to be significant, indicating a difference in the means, and rejecting the null hypothesis.

Referring to our working hypothesis, that the treatment of using robots will influence TUG-K scores, we perform a t-test for independent samples. The scores are from students in the same course so that we can assume equal variance of the scores. Levene's Test for variance p-value is reported as 0.472, so this affirms the equality of variance. The significance of the t-test is too high to reject the null hypothesis, so it will be assumed that there was not a significant difference in the TUG-K scores between the control and treatment groups.

**Independent Samples Test**

| | | Levene's Test for Equality of Variances | | t-test for Equality of Means | | | |
|---|---|---|---|---|---|---|---|
| | | F | Sig. | t | df | Sig. (2-tailed) | Mean Difference |
| DeltaTUGK | Equal variances assumed | .533 | .472 | -1.590 | 27 | .123 | -3.27381 |
| | Equal variances not assumed | | | -1.854 | 17.971 | .080 | -3.27381 |

| Treatment group | Unstandardized Coefficients B | Std. Error | Standardized Coefficients Beta | t | Sig. |
|---|---|---|---|---|---|
| (Constant) | -.325 | 4.877 | | -.067 | .958 |
| OverallFav | 1.220 | .514 | 3.852 | 2.376 | .254 |
| OverAllUnfav | 1.089 | .487 | 2.871 | 2.238 | .268 |
| AllcategoriesFav | -.484 | .531 | -1.914 | -.910 | .530 |
| PersonalInterestFav | -.432 | .207 | -2.737 | -2.092 | .284 |
| PersonalInterestUnfav | -.376 | .330 | -1.746 | -1.139 | .459 |
| RealWorldConnectionFav | -.072 | .121 | -.466 | -.592 | .660 |
| RealWorldConnectionUnfav | .024 | .167 | .112 | .145 | .908 |
| PSGeneralFav | -.163 | .435 | -.807 | -.376 | .771 |
| PSGeneralUnfav | -1.280 | .561 | -5.470 | -2.281 | .263 |
| PSConfidenceFav | .064 | .353 | .348 | .180 | .887 |
| PSConfidenceUnfav | .575 | .277 | 3.349 | 2.074 | .286 |
| PSSophisticationFav | .102 | .298 | .395 | .343 | .790 |
| PSSophisticationUnfav | .503 | .521 | 2.203 | .966 | .511 |
| SensesMakingEffortFav | .390 | .246 | 1.927 | 1.585 | .358 |
| SensesMakingEffortUnfav | .293 | .328 | .844 | .892 | .536 |
| ConceptUnderstandingFav | .314 | .185 | 1.639 | 1.697 | .339 |
| ConceptUnderstandingUnfav | .523 | .242 | 2.579 | 2.165 | .275 |
| AppConcepUnderstandFav | -1.016 | .420 | -3.626 | -2.418 | .250 |
| AppConcepUnderstanUnfav | -1.408 | .478 | -6.260 | -2.944 | .208 |

This regression seemed to fit well with an $R^2$ of 0.822, or 82% of the variability in the TUG-K scores was measured against the CLASS. When looking at the standardized Coefficients from the linear regression, we see that the overall favorability, *OverallFav,* and *OverAllUnfav*, has a reasonably significant weight (2.7 and 2.1); however, only the *OverallFav* variable seems to be significant. The problem solving general favorability variable *PSGeneralFav* has the largest beta weight, indicated it is the most important predictor of the change in the TUG-K score. It also seems to be the most significant weight with p=0.03. Also, as might be expected, the variables connected to confidence (*PSConfidenceFav, Unfav*) played a significant role and had a moderate beta value. It is somewhat surprising that the Real-World connection (*RealWorldConnectionFav*, *Unfav*) and Sensemaking variables the were not significant. According to the developers of the CLASS, the name of the category or variable does not fully describe that category. The variable is defined as *PSGeneralFav* contains the following statements*:*

> I do not expect physics equations to help my understanding of the ideas; they are just for doing calculations.

> If I get stuck on a physics problem on my first try, I usually try to figure out a different way that works.

> Nearly everyone is capable of understanding physics if they work at it.

> I enjoy solving physics problems.

Therefore, we can say with a reasonable amount of confidence that favorability toward problem-solving enhances the score on the TUG-K`

**Coefficients – Dependent Variable is delta TUGK**

| | Unstandardized Coefficients | | Standardized Coefficients | | |
|---|---|---|---|---|---|
| All sections | B | Std. Error | Beta | t | Sig. |
| (Constant) | 1.260 | 2.744 | | .459 | .660 |
| §OverallFav | .729 | .317 | 2.716 | 2.299 | .055§ |
| OverAllUnfav | .696 | .396 | 2.118 | 1.759 | .122 |
| AllcategoriesFav | .621 | .952 | 2.727 | .652 | .535 |
| AllcategoriesUnfav | -.332 | 1.925 | -1.112 | -.172 | .868 |
| PersonalInterestFav | -.066 | .152 | -.418 | -.432 | .679 |
| PersonalInterestUnfav | .128 | .245 | .624 | .524 | .617 |
| RealWorldConnectionFav | -.114 | .148 | -.736 | -.771 | .466 |
| RealWorldConnectionUnfav | -.001 | .244 | -.006 | -.006 | .996 |
| §PSGeneralFav | -.785 | .289 | -4.312 | -2.714 | .030§ |
| §PSGeneralUnfav | -.868 | .400 | -4.032 | -2.169 | .067§ |
| §PSConfidenceFav | .278 | .126 | 1.752 | 2.198 | .064§ |
| §PSConfidenceUnfav | .401 | .140 | 2.390 | 2.868 | .024§ |
| PSSophisticationFav | -.172 | .190 | -.726 | -.907 | .395 |
| PSSophisticationUnfav | -.035 | .238 | -.189 | -.148 | .886 |
| SensesMakingEffortFav | -.130 | .208 | -.729 | -.628 | .550 |
| SensesMakingEffortUnfav | -.043 | .400 | -.139 | -.109 | .917 |
| ConceptUnderstandingFav | .101 | .120 | .526 | .840 | .429 |
| ConceptUnderstandingUnfav | .336 | .203 | 1.804 | 1.653 | .142 |
| §AppConcepUnderstandFav | -.487 | .255 | -1.716 | -1.911 | .098§ |
| AppConcepUnderstanUnfav | -.405 | .456 | -2.214 | -.887 | .404 |

When the CLASS was initially developed, K.A. Douglas (Douglas, Yale, Bennett, Haugan, & Bryan, 2014) found the following components

Factor 1: Personal Application and Relation to Real World

Factor 2: Problem Solving and Learning

Factor 3: Effort and Sense-Making

Earlier use of the CLASS has identified several sets of variables. While these variables were not initially used to construct the instrument, latent variables were considered during its development. The CLASS was developed with a set of predetermined latent variables W.K. Adams et al. found the following emergent factors. The term used in the paper was Category.

Category 1 SS$_a$ Real-world connection and personal interest
Category 2 SS$_a$ Real-world connection and personal interest
Category 3 BQ Conceptual understanding
Category 6 SS Sensemaking/effort

The boundaries between these factors are approximate, and several of them will overlap, depending upon the population being studied. "There is no such thing as a perfect category or factor." W.K. Adams

From our data, four components were extracted and identified. In the same manner, as used in other studies, the variable with the highest coefficient is used to identify the latent factor.

1 – Problems solving sophistication, real-world connection & personal interest

2 – Sensemaking, effort, and personal interest

3 – problem-solving confidence

4 - conceptual understanding

**Component Matrix**

|  | Rescaled Component | | | |
|---|---|---|---|---|
|  | 1 | 2 | 3 | 4 |
| PersonalInterestFav | .737 | .496 | -.206 |  |
| PersonalInterestUnfav | -.605 | .382 | .573 | .147 |
| RealWorldConnectionFav | .782 | .420 | -.185 | .243 |
| RealWorldConnectionUnfav | -.456 | .327 | .643 | -.161 |
| PSGeneralFav | .928 | .171 | .147 | -.179 |
| PSGeneralUnfav | -.575 | .685 | .313 |  |
| PSConfidenceFav | .822 |  | .394 | -.314 |
| PSConfidenceUnfav | -.493 | .706 |  | .316 |
| PSSophisticationFav | .885 |  | .249 | -.136 |
| PSSophisticationUnfav | -.336 | .862 |  | .138 |
| SensesMakingEffortFav | .690 | .556 | .120 | .195 |
| SensesMakingEffortUnfav | -.107 | .385 | .449 | -.300 |
| ConceptUnderstandingFav | .755 |  | .168 | .470 |
| ConceptUnderstandingUnfav | -.173 | .829 | -.174 | -.446 |
| AppConcepUnderstandFav | .751 |  | .273 | .396 |
| AppConcepUnderstanUnfav |  | .880 | -.252 | -.187 |
| DeltaTUGK |  |  | -.259 | .140 |

Extraction Method: Principal Component Analysis.

**Component Standardized Score Coefficient Matrix**

|  | Component | | | |
|---|---|---|---|---|
|  | 1 | 2 | 3 | 4 |
| PersonalInterestFav | .137 | -.117 | .261 | -.119 |
| PersonalInterestUnfav | .042 | .262 | -.168 | .257 |
| RealWorldConnectionFav | .269 | .133 | .009 | -.346 |
| RealWorldConnectionUnfav | -.005 | .094 | -.057 | .346 |
| PSGeneralFav | .115 | -.118 | .061 | .203 |
| PSGeneralUnfav | .026 | .195 | -.028 | .132 |
| PSConfidenceFav | .103 | -.195 | .022 | .583 |
| PSConfidenceUnfav | .099 | .403 | -.075 | -.122 |
| PSSophisticationFav | .071 | -.049 | -.014 | .145 |
| PSSophisticationUnfav | .050 | .197 | .104 | -.101 |
| SensesMakingEffortFav | .222 | .177 | -.057 | -.009 |
| SensesMakingEffortUnfav | -.005 | -.002 | .021 | .164 |
| ConceptUnderstandingFav | .229 | .231 | -.292 | -.128 |
| ConceptUnderstandingUnfav | -.098 | -.178 | .421 | .138 |
| AppConcepUnderstandFav | .102 | .100 | -.132 | -.012 |
| AppConcepUnderstanUnfav | -.021 | -.050 | .339 | -.048 |
| DeltaTUGK | .000 | .001 | .001 | -.008 |

Extraction Method: Principal Component Analysis.

Rotation Method: Varimax with Kaiser Normalization.

Component Scores.

Coefficients are standardized.

## Discussion

As was stated in the introduction, there are two research questions that were being addressed. 1) Will a student-centered, technology-based activity influence the ability of novice physics students to interpret kinematic graphs? More importantly 2) Does attitude about learning differ between physics students and non- science students

As was mentioned before, the treatment had no effect on the TUG - K scores. We have seen as a result of our analysis and study that the answer to research question number one is no; the student-centered robot activity that we presented to the non-science majors had no effect on their understanding of graphing. There was no significant difference between the control group and treatment group scores on the TUG-K. While a plethora of physics education research tells us there should be a difference in graphing ability, in this case, the treatment was too little and too short. The participants were exposed to real-time graphing for only half an hour out of 30 hours of lab work during the semester. The subjects only drew two graphs with the equipment, allowing little time to absorb the consequences of the activity. The activity for this study was inserted into a regular semester, and there was not much available time in the course to dedicate to this activity. In the future, more time would be spent on treatment activity.

As far as the other research question is concerned, there is an indication that for the non-science student, attitude does play a role in interpreting graphs. Eighty percent of the variability in the TUG-K scores was related to the CLASS score. When looking at the standardized Coefficients from the linear regression, we see that the overall favorability, *OverallFav,* and *OverAllUnfav*, has a reasonably significant weight (2.7 and 2.1); however, only the *OverallFav* variable seems to be significant.

# Appendix A

# Results from the Colorado Learning About Science Survey (CLASS)

| | PRE | | | | POST | | | SHIFT | | |
|---|---|---|---|---|---|---|---|---|---|---|
| Agree | Neutral | Disagree | | Agree | Neutral | Disagree | Agree | Disagree | Net Shift | |
| PERSONAL INTEREST: DO STUDENTS FEEL A PERSONAL INTEREST IN /CONNECTION TO PHYSICS ||||||||||||
| 3. I think about the physics I experience in everyday life. ||||||||||||
| 58% | 21% | 21% | | 42% | 18% | 39% | -15% | 18% | NOVICE |
| 11. I am not satisfied until I understand why something works the way it does. ||||||||||||
| 88% | 3% | 9% | | 67% | 27% | 6% | -21% | -3% | NOVICE |
| 14. I study physics to learn knowledge that will be useful in my life outside of school. ||||||||||||
| 61% | 21% | 18% | | 33% | 39% | 27% | -27% | 9% | NOVICE |
| 25. I enjoy solving physics problems. ||||||||||||
| 70% | 27% | 3% | | 36% | 27% | 36% | -33% | 33% | NOVICE |
| 28. Learning physics changes my ideas about how the world works. ||||||||||||
| 30% | 15% | 55% | | 58% | 30% | 12% | 27% | -42% | EXPERT |
| 30. Reasoning skills used to understand physics can be helpful to me in my everyday life. ||||||||||||
| 76% | 15% | 9% | | 61% | 33% | 6% | -15% | -3% | NOVICE |
| REAL WORLD CONNECTION: SEEING THE CONNECTION BETWEEN PHYSICS AND REAL LIFE ||||||||||||
| 28. Learning physics changes my ideas about how the world works. ||||||||||||
| 30% | 15% | 55% | | 58% | 30% | 12% | 27% | -42% | EXPERT |
| 30. Reasoning skills used to understand physics can be helpful to me in my everyday life. ||||||||||||
| 76% | 15% | 9% | | 61% | 33% | 6% | -15% | -3% | NOVICE |
| colspan="10" | 35. The subject of physics has little relation to what I experience in the real world. |
| 45% | 30% | 24% | | 27% | 42% | 30% | -18% | 6% | EXPERT |
| colspan="10" | 37. To understand physics, I sometimes think about my personal experiences and relate them to the topic being analyzed. |
| 30% | 27% | 42% | | 45% | 24% | 30% | 15% | -12% | EXPERT |
| colspan="10" | PROBLEM SOLVING GENERAL: |
| colspan="10" | 13. I do not expect physics equations to help my understanding of the ideas; they are just for doing calculations. |
| 30% | 24% | 45% | | 36% | 33% | 30% | 6% | -15% | NOVICE |
| colspan="10" | 15. If I get stuck on a physics problem on my first try, I usually try to figure out a different way that works. |
| 48% | 21% | 30% | | 64% | 24% | 12% | 15% | -18% | EXPERT |
| colspan="10" | 16. Nearly everyone is capable of understanding physics if they work at it. |
| 42% | 33% | 24% | | 52% | 24% | 24% | 9% | 0% | |
| colspan="10" | 25. I enjoy solving physics problems. |
| 70% | 27% | 3% | | 36% | 27% | 36% | -33% | 33% | NOVICE |
| colspan="10" | 26. In physics, mathematical formulas express meaningful relationships among measurable quantities. |
| 58% | 27% | 15% | | 64% | 30% | 6% | 6% | -9% | EXPERT |
| colspan="10" | 34. I can usually figure out a way to solve physics problems. |
| 36% | 24% | 39% | | 45% | 33% | 21% | 9% | -18% | EXPERT |
| colspan="10" | 40. If I get stuck on a physics problem, there is no chance I'll figure it out on my own. |
| 45% | 33% | 21% | | 39% | 24% | 36% | -6% | 15% | EXPERT |
| colspan="10" | PROBLEM SOLVING CONFIDENCE |
| colspan="10" | 15. If I get stuck on a physics problem on my first try, I usually try to figure out a different way that works. |
| 48% | 21% | 30% | | 64% | 24% | 12% | 15% | -18% | EXPERT |
| colspan="10" | 16. Nearly everyone is capable of understanding physics if they work at it. |
| 42% | 33% | 24% | | 52% | 24% | 24% | 9% | 0% | |
| colspan="10" | 34. I can usually figure out a way to solve physics problems. |
| 36% | 24% | 39% | | 45% | 33% | 21% | 9% | -18% | EXPERT |
| colspan="10" | 40. If I get stuck on a physics problem, there is no chance I'll figure it out on my own. |
| 45% | 33% | 21% | | 39% | 24% | 36% | -6% | 15% | EXPERT |
| colspan="10" | PROBLEM SOLVING SOPHISTOCATION |
| colspan="10" | 5. After I study a topic in physics and feel that I understand it, I have difficulty solving problems on the same topic. |
| 39% | 33% | 27% | | 39% | 45% | 15% | 0% | -12% | NOVICE |
| colspan="10" | 21. If I don't remember an equation needed to solve a problem on an exam, there's nothing much I can do (legally!) to come up with it. |
| 52% | 27% | 21% | | 42% | 45% | 12% | - | -9% | |

| | | | | | | | 9% | | |
|---|---|---|---|---|---|---|---|---|---|
| 25. I enjoy solving physics problems. | | | | | | | | | |
| 70% | 27% | 3% | | 36% | 27% | 36% | -33% | 33% | NOVICE |
| 34. I can usually figure out a way to solve physics problems. | | | | | | | | | |
| 36% | 24% | 39% | | 45% | 33% | 21% | 9% | -18% | EXPERT |
| 40. If I get stuck on a physics problem, there is no chance I'll figure it out on my own. | | | | | | | | | |
| 45% | 33% | 21% | | 39% | 24% | 36% | -6% | 15% | EXPERT |
| SENSEMAKING/EFFORT: FOR ME (THE STUDENT) EXERTING THE EFFORT NEEDED TOWARDS SENSE-MAKING IS WORTHWHILE | | | | | | | | | |
| 11. I am not satisfied until I understand why something works the way it does. | | | | | | | | | |
| 88% | 3% | 9% | | 67% | 27% | 6% | -21% | -3% | NOVICE |
| 32. Spending a lot of time understanding where formulas come from is a waste of time. | | | | | | | | | |
| 58% | 21% | 21% | | 39% | 27% | 33% | -18% | 12% | EXPERT |
| 36. There are times I solve a physics problem more than one way to help my understanding. | | | | | | | | | |
| 52% | 15% | 33% | | 52% | 42% | 6% | 0% | -27% | EXPERT |
| 39. When I solve a physics problem, I explicitly think about which physics ideas apply to the problem. | | | | | | | | | |
| 33% | 27% | 39% | | 61% | 27% | 12% | 27% | -27% | EXPERT |
| CONCEPTUAL UNDERSTANDING: UNDERSTANDING THAT PHYSICS IS COHERENT AND IS ABOUT MAKING-SENSE, DRAWING CONNECTIONS, AND REASONING NOT MEMORIZING. MAKING SENSE OF MATH | | | | | | | | | |
| 1. A significant problem in learning physics is being able to memorize all the information I need to know. | | | | | | | | | |
| 82% | 12% | 6% | | 45% | 21% | 33% | -36% | 27% | EXPERT |
| 5. After I study a topic in physics and feel that I understand it, I have difficulty solving problems on the same topic. | | | | | | | | | |
| 39% | 33% | 27% | | 39% | 45% | 15% | 0% | -12% | NOVICE |
| 6. Knowledge in physics consists of many disconnected topics. | | | | | | | | | |
| 15% | 58% | 27% | | 39% | 36% | 24% | 24% | -3% | NOVICE |
| 13. I do not expect physics equations to help my understanding of the ideas; they are just for doing calculations. | | | | | | | | | |
| 30% | 24% | 45% | | 36% | 33% | 30% | 6% | -15% | NOVICE |
| 21. If I don't remember an equation needed to solve a problem on an exam, there's nothing much I can do (legally!) to come up with it. | | | | | | | | | |
| 52% | 27% | 21% | | 42% | 45% | 12% | -9% | -9% | |
| 32. Spending a lot of time understanding where formulas come from is a waste of time. | | | | | | | | | |
| 58% | 21% | 21% | | 39% | 27% | 33% | -18 | 12% | EXPERT |

| | | | | | | | % | |
|---|---|---|---|---|---|---|---|---|
| APPLIED CONCEPTUAL UNDERSTANDING: UNDERSTANDING AND APPLYING A CONCEPTUAL APPROACH AND REASONING IN PROBLEM SOLVING, NOT MEMORIZING OR FOLLOWING PROBLEM-SOLVING RECIPES ||||||||||
| 1. A significant problem in learning physics is being able to memorize all the information I need to know. ||||||||||
| 82% | 12% | 6% | | 45% | 21% | 33% | -36% | 27% | EXPERT |
| 5. After I study a topic in physics and feel that I understand it, I have difficulty solving problems on the same topic. ||||||||||
| 39% | 33% | 27% | | 39% | 45% | 15% | 0% | -12% | NOVICE |
| 6. Knowledge in physics consists of many disconnected topics. ||||||||||
| 15% | 58% | 27% | | 39% | 36% | 24% | 24% | -3% | NOVICE |
| 8. When I solve a physics problem, I locate an equation that uses the variables given in the problem and plug in the values. ||||||||||
| 42% | 33% | 24% | | 79% | 18% | 3% | 36% | -21% | NOVICE |
| 21. If I don't remember an equation needed to solve a problem on an exam, there's nothing much I can do (legally!) to come up with it. ||||||||||
| 52% | 27% | 21% | | 42% | 45% | 12% | -9% | -9% | |
| 40. If I get stuck on a physics problem, there is no chance I'll figure it out on my own ||||||||||
| 45% | 33% | 21% | | 39% | 24% | 36% | -6% | 15% | EXPERT |
| REMAINING STATEMENTS FOR WHICH THERE IS A CONSISTENT EXPERT PERSPECTIVE ||||||||||
| 2. When I am solving a physics problem, I try to decide what would be a reasonable value for the answer. ||||||||||
| 24% | 15% | 61% | | 79% | 9% | 12% | 55% | -48% | EXPERT |
| 10. There is usually only one correct approach to solving a physics problem. ||||||||||
| 58% | 21% | 21% | | 30% | 30% | 39% | -27% | 18% | EXPERT |
| 12. I cannot learn physics if the teacher does not explain things well in class. ||||||||||
| 42% | 24% | 33% | | 82% | 12% | 6% | 39% | -27% | NOVICE |
| 17. Understanding physics basically means being able to recall something you've read or been shown. ||||||||||
| 39% | 27% | 33% | | 27% | 42% | 30% | -12% | -3% | |
| 18. There could be two different correct values for the answer to a physics problem if I use two different approaches. ||||||||||
| 67% | 6% | 27% | | 33% | 39% | 27% | -33% | 0% | EXPERT |
| 19. To understand physics I discuss it with friends and other students. ||||||||||
| 48% | 21% | 30% | | 70% | 18% | 12% | 21% | -18% | EXPERT |
| 20. I do not spend more than five minutes stuck on a physics problem before giving up or seeking help from someone else. ||||||||||
| 52% | 30% | 18% | | 42% | 33% | 24% | -9% | 6% | EXPERT |

| 29. To learn physics, I only need to memorize solutions to sample problems. |||||||||
|---|---|---|---|---|---|---|---|---|
| 58% | 36% | 6% | | 30% | 27% | 42% | -27% | 36% | EXPERT |

| 38. It is possible to explain physics ideas without mathematical formulas. |||||||||
|---|---|---|---|---|---|---|---|---|
| 52% | 30% | 18% | | 30% | 48% | 21% | -21% | 3% | NOVICE |

### LEARNING STYLE Q's (not a validated category): WHAT STUDENTS BELIEVE TO BE USEFUL FOR LEARNING

| 4. It is useful for me to do lots and lots of problems when learning physics. |||||||||
|---|---|---|---|---|---|---|---|---|
| 39% | 30% | 30% | | 58% | 36% | 6% | 18% | -24% | |

| 9. I find that reading the text in detail is a good way for me to learn physics. |||||||||
|---|---|---|---|---|---|---|---|---|
| 45% | 21% | 33% | | 39% | 33% | 27% | -6% | -6% | |

| 12. I cannot learn physics if the teacher does not explain things well in class. |||||||||
|---|---|---|---|---|---|---|---|---|
| 42% | 24% | 33% | | 82% | 12% | 6% | 39% | -27% | NOVICE |

| 16. Nearly everyone is capable of understanding physics if they work at it. |||||||||
|---|---|---|---|---|---|---|---|---|
| 42% | 33% | 24% | | 52% | 24% | 24% | 9% | 0% | |

| 19. To understand physics I discuss it with friends and other students. |||||||||
|---|---|---|---|---|---|---|---|---|
| 48% | 21% | 30% | | 70% | 18% | 12% | 21% | -18% | EXPERT |

| 33. I find carefully analyzing only a few problems in detail is a good way for me to learn physics. |||||||||
|---|---|---|---|---|---|---|---|---|
| 39% | 42% | 18% | | 55% | 27% | 18% | 15% | 0% | |

### STATEMENTS ON SLATE FOR REVISION

| 7. As physicists learn more, most physics ideas we use today are likely to be proven wrong. |||||||||
|---|---|---|---|---|---|---|---|---|
| 73% | 15% | 12% | | 33% | 48% | 18% | -39% | 6% | EXPERT |

| 41. It is possible for physicists to carefully perform the same experiment and get two very different results that are both correct. |||||||||
|---|---|---|---|---|---|---|---|---|
| 45% | 36% | 18% | | 61% | 24% | 15% | 15% | -3% | NOVICE |

Robot activity

In this lesson, we are going to examine motion graphs again. We have a small green "robot" called the scribbler 3. You are going to draw graphs on the computer. You will send that graph to the robot and the robot will move according to the graph you draw.

**Instructions:**

1) Read the first scenario.
2) For that scenario, sketch on paper, the graph corresponding to the motion depicted in this scenario.
3) On the desk top of the computer, there is a red icon to the software s2mmsKinematicGUI
4) Click on the icon to start s2mmKintaticsGUI. The program opens with a simple graph. Click on the vertical, (Y) axis, on the left. The axis will turn to red and a select box will appear at the bottom (See figure to the right)
5) Set the range from 0.0 to 90.0 cm and the graph increments (by) to 10cm. (This has no bearing on the results, it just makes the graph easier to read.

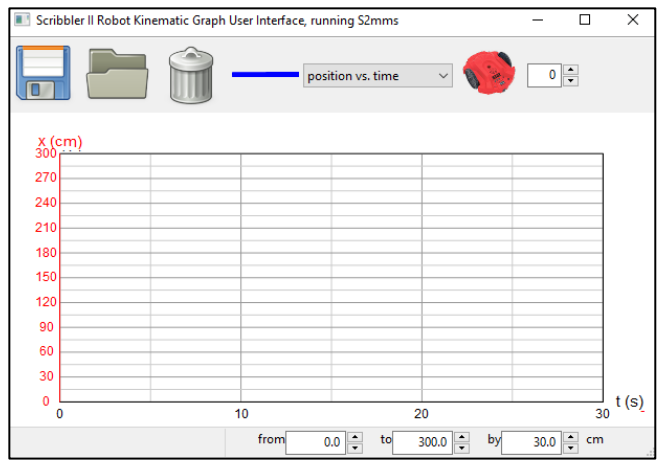

6) Be sure the USB cable connects the computer to the laptop.
7) On the s2mm graph, draw the predicted graph you predicted on paper. Click on the blue line at the top of the graph, then click on the graph itself. You can click and drag on the line segments to change their slope. If you would like to make a curved line, click in the middle of the line segment, and drag the center of the segment up or down to change the line.
8) When you are satisfied with the line, click on the red icon at the top left. That will send the graph to the robot.
9) Press the **blue** button on the robot, and it should move according to the graph you drew.
10) Lastly, answer the question – Does the robot move as you predicted? If not explain how it was different.
11) Repeat these steps for the other scenario.

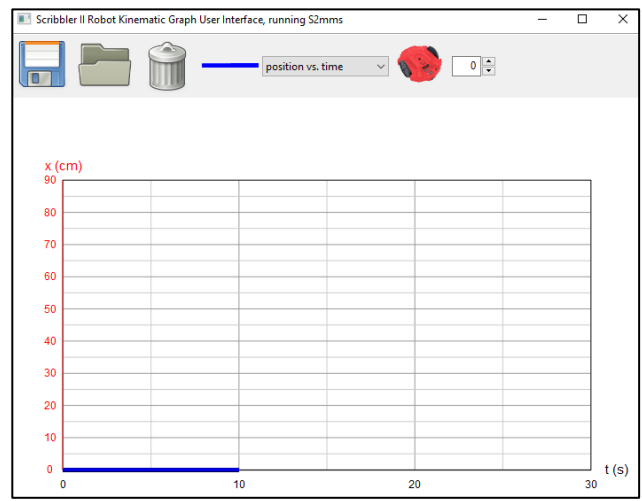